\documentclass[conference]{IEEEtran}
\IEEEoverridecommandlockouts
\usepackage{cite}
\usepackage{amsmath,amssymb,amsfonts}
\usepackage{algorithmic}
\usepackage{graphicx}
\usepackage{textcomp}
\usepackage{xcolor}
\usepackage{hyperref}
\def\BibTeX{{\rm B\kern-.05em{\sc i\kern-.025em b}\kern-.08em
    T\kern-.1667em\lower.7ex\hbox{E}\kern-.125emX}}
\begin{document}

\title{Spatio-Temporal 3D Point Clouds from WiFi-CSI Data via Transformer Networks 
\thanks{This research was supported by the Research Council of Finland 6G Flagship Programme (Grant Number: 346208), Horizon Europe CONVERGE project (Grant 101094831), and Business Finland WISEC project (Grant 3630/31/2024).}
}

\author{\IEEEauthorblockN{Tuomas Määttä, Sasan Sharifipour, Miguel Bordallo López, Constantino Álvarez Casado}
\IEEEauthorblockA{\textit{Center for Machine Vision and Signal Analysis (CMVS)} \\
\textit{University of Oulu}, Finland \\
\{tuomas.maata, sasan.sharifipour, miguel.bordallo, constantino.alvarezcasado\}@oulu.fi}
}

\maketitle

\begin{abstract}
Joint communication and sensing (JC\&S) is emerging as a key component in 5G and 6G networks, enabling dynamic adaptation to environmental changes and enhancing contextual awareness for optimized communication. By leveraging real-time environmental data, JC\&S improves resource allocation, reduces latency, and enhances power efficiency, while also supporting simulations and predictive modeling. This makes it a key technology for reactive systems and digital twins. These systems can respond to environmental events in real-time, offering transformative potential in sectors like smart cities, healthcare, and Industry 5.0, where adaptive and multimodal interaction is critical to enhance real-time decision-making. In this work, we present a transformer-based architecture that processes temporal Channel State Information (CSI) data, specifically amplitude and phase, to generate 3D point clouds of indoor environments. The model utilizes a multi-head attention to capture complex spatio-temporal relationships in CSI data and is adaptable to different CSI configurations. We evaluate the architecture on the MM-Fi dataset, using two different protocols to capture human presence in indoor environments. The system demonstrates strong potential for accurate 3D reconstructions and effectively distinguishes between close and distant objects, advancing JC\&S applications for spatial sensing in future wireless networks. The code is available at: \url{https://github.com/Arritmic/csi2pointcloud}.

\end{abstract}


%
%
%
%

\section{Introduction}
The convergence of joint communication and sensing (JC\&S) is becoming a foundational element of future wireless networks, particularly in the context of 5G and 6G technologies. As the demand for higher data rates, ultra-low latency, and pervasive connectivity increases, integrating sensing capabilities into communication networks allows for greater contextual awareness. This enhances network performance by optimizing communication based on environmental conditions, user behavior, and spatial awareness. Such contextual intelligence is crucial for real-time decision-making, providing opportunities to optimize resource allocation, interference management, and power efficiency. The ability to sense the environment and respond to changes in real-time enables the development of event-driven systems, which are fundamental for reactive systems and digital twins. These systems replicate physical environments and adapt to dynamic changes, allowing for advanced applications in smart cities, healthcare monitoring, and other fields \cite{susarla2024contactless}. By incorporating multimodal sensing, including radio frequency (RF) data and other technologies, future networks will create more detailed and accurate representations of their environments.

\begin{figure}[ht!]
 \begin{center}
   \includegraphics*[width=0.49\textwidth]{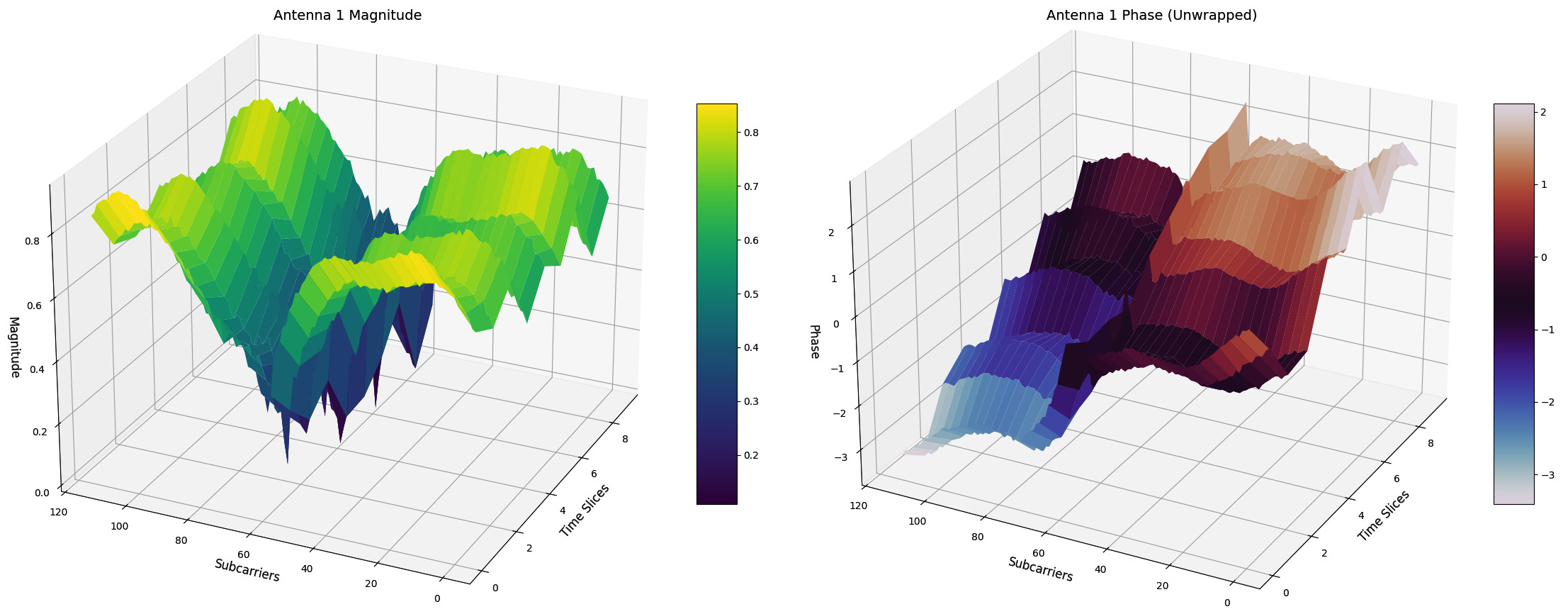}
 \end{center}
 \caption{Magnitude (left) and unwrapped phase (right) of CSI data from a receiver antenna using the 802.11n WiFi standard at 5GHz with 40MHz bandwidth, across 114 subcarriers and 10 time slices. The plots illustrate signal strength variations and phase shifts influenced by environmental factors, critical for spatial and temporal channel analysis.}
 \label{fig:csidata}
\end{figure}

In this work, we propose a novel deep learning (DL) architecture that uses CSI (Channel State Information) amplitude and phase data, as shown in Figure \ref{fig:csidata}, to generate 3D point cloud views of indoor environments. The architecture is designed to handle temporal CSI data and leverages transformers with multi-head attention to capture the complex spatiotemporal relationships between the transmitted and received signals. Our approach enables the system to generate 3D reconstructions of environments, distinguishing between close objects (such as humans) and far objects (such as room structures).

We validate our architecture using the MM-Fi Database \cite{yang2024mmfi}, which captures human presence and motion in indoor environments. This dataset enable the proposed model to learn how to transform temporal CSI data (amplitude and phase) into 3D point cloud representations, demonstrating the potential for joint communication and sensing in creating context-aware, multimodal systems that can adapt to and react to environmental changes in real-time.

%
%
%
%

\section{Related Work}
The use of ambient sensors, including cameras, LiDAR, radar, and radio frequency (RF)-based sensors, has gained significant attention in recent years for sensing and modeling environments. These sensors provide the capability to gather detailed information about the context in which systems operate, offering valuable insights for applications such as indoor localization, object detection, and human activity recognition \cite{susarla2024contactless}. The ability to reconstruct and understand environments plays a crucial role in fields such as autonomous navigation, healthcare, and smart city infrastructure \cite{tang2022perception}. Digital representations of environments, including digital twins, have emerged as important tools in both industrial and research contexts \cite{alkhateeb2023real}. Digital twins refer to the virtual replication of physical systems or spaces, enabling real-time monitoring and predictive analysis. In addition to improving operational efficiency, digital twins facilitate remote diagnostics and provide a foundation for real-time simulations \cite{alkhateeb2023real}. In fields such as manufacturing, urban planning, and healthcare, digital twins offer practical benefits, enhancing decision-making processes and improving system control \cite{piras2024digital}. The construction of accurate 3D models from sensor data is central to the realization of these digital twin systems.

Several methods for generating 3D models rely on depth and stereo vision cameras. Depth cameras, like Kinect, use infrared light to measure depth and are commonly applied in indoor mapping, robotics, and human-computer interaction, producing point clouds and meshes in real time \cite{F20143}. Stereo cameras, which estimate depth by comparing offset images, are often used in autonomous systems for object detection. While Kinect provides better distance precision, stereo cameras perform well at shorter ranges \cite{Julian2021comparison}. Despite improvements, challenges remain in depth data quality and real-time reconstruction, especially for consumer sensors \cite{Jianwei2022High}. Systems combining multiple Kinect sensors have advanced real-time 3D reconstructions of moving objects, such as humans, in dynamic settings \cite{D2013Real}.


Radar and mmWave radar systems are used to generate 3D models by measuring radio wave reflections, allowing object detection and distance estimation, particularly in environments where optical systems struggle, such as low-light or obstructed conditions. In the automotive industry, mmWave radar is widely used for collision avoidance and autonomous driving by generating 3D representations of surroundings. It is also applied in robotics and human sensing for real-time movement tracking and posture estimation. Deep learning has been increasingly employed to enhance 3D point cloud quality from mmWave radar data. Models like R2P generate dense and accurate 3D point clouds from sparse radar data, outperforming approaches like PointNet \cite{Sun2022R2P}. Other methods, such as radar-based perception for mobile robots, improve point cloud quality for robotics \cite{Cheng2022Novel}. DeepPoint uses conditional GANs to create smooth 3D point clouds from 2D depth images derived from mmWave radar data \cite{Yue2021DeepPoint}. Recent approaches, such as mmWaveNet, directly generate high-quality point clouds from radar reflections without intermediate transformations \cite{Gu2023Poster}. In automotive applications, the MILLIPOINT system provides low-cost 3D point cloud generation using vehicle radars and synthetic aperture radar (SAR) imaging \cite{Qian20203D}. These methods underscore radar’s increasing role in producing robust 3D models for both dynamic and static environments.

Recent work has also explored the use of Wi-Fi signals, particularly CSI data, to sense human movement and reconstruct environments. CSI captures how the Wi-Fi signal interacts with the physical environment, providing information about the changes in amplitude and phase across different subcarriers. This data can be leveraged to detect human presence, and track movement  \cite{CSI_F2024}. It can also be used to generate spatial models of indoor environments. Wi-Fi-based sensing is non-invasive and benefits from being part of the existing communication infrastructure, making it an attractive alternative to more expensive and power-intensive sensors. However, while Wi-Fi-based methods show promise, they are still in the early stages of development, and challenges remain in terms of accuracy and spatial resolution. The combination of multiple sensing modalities has been proposed to improve environmental sensing by leveraging the strengths of each sensor to increase accuracy and reliability. For instance, integrating camera data with RF or radar signals has been suggested as a way to combine detailed visual information with the ability of RF to penetrate obstacles. This approach is particularly valuable for applications requiring high precision and adaptability to dynamic environments \cite{alkhateeb2023real}.


Despite significant advancements in sensing technologies, the use of CSI amplitude and phase information, specifically for generating 3D point cloud models remains relatively unexplored. Most of the existing work focuses on human activity recognition or indoor localization using CSI, without extending the data to full 3D spatial reconstructions. In the next section, we present a novel method that processes CSI data with transformer-based networks to generate 3D point clouds. This approach bridges the gap between communication-based sensing and detailed environmental modeling, demonstrating the potential for joint communication and sensing in 5G and 6G networks.

%
%
%
%
\section{Methodology}

In this section, we present the CSI2PointCloudModel (CSI2PC) architecture, a transformer-based framework to process spatio-temporal CSI (Channel State Information) data and generate 3D point clouds, as depicted in Figure \ref{fig:architecture}. CSI data reflects how WiFi signals propagate and interact with the environment. By leveraging the spatio-temporal nature of CSI data, our architecture learns to reconstruct 3D spatial structures, thus making it a valuable tool for indoor environment sensing.


\subsection{Proposed CSI2PC architecture}

The input to the model is CSI data collected over time, structured as a multi-dimensional tensor. Each element in the tensor corresponds to the amplitude and phase information from WiFi signals, organized by the number of antennas, subcarriers, and time slices. This data is represented as $ x \in \mathbb{R}^{A \times S \times 2 \times T} $, where $A$ is the number of antennas, $S$ is the number of subcarriers, and $T$ is the number of time slices. The two channels represent amplitude and phase. To prepare the data for processing, the antenna and subcarrier dimensions are merged into a single feature dimension. This results in a tensor of shape $ x \in \mathbb{R}^{F \times 2 \times T} $, where $ F = A \times S $ represents the total number of antenna-subcarrier pairs. This transformation aligns the temporal data for each antenna-subcarrier pair, facilitating efficient processing in the subsequent layers of the model.

\begin{figure*}[ht!]
\vspace{-1mm}
 \begin{center}
   \includegraphics*[width=0.99\textwidth]{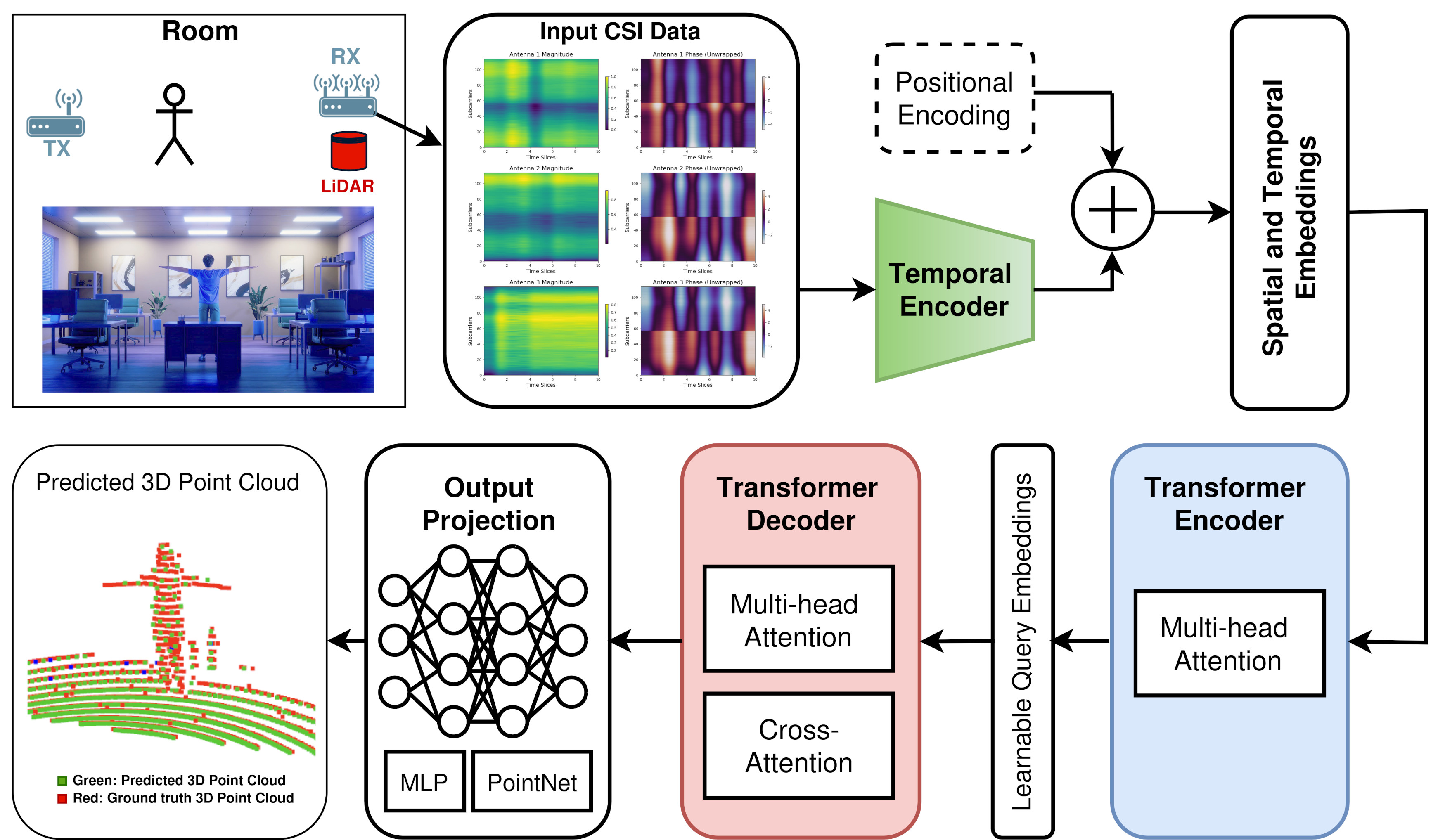}
 \end{center}
 \caption{Architecture of the CSI2PointCloud Model, from input CSI data to 3D point cloud output, highlighting key stages of data transformation and encoding through transformer layers.}
 \label{fig:architecture}
\end{figure*}

Our architecture consists of four stages, as shown in Figure \ref{fig:architecture}. First, temporal encoding uses a convolutional layer to extract patterns over time, capturing amplitude and phase variations related to environmental changes. Positional encoding adds learned embeddings for antennas and subcarriers to preserve spatial context. The Transformer Encoder then applies multi-head attention to model dependencies between antenna-subcarrier pairs, capturing both spatial and temporal interactions. Finally, the Transformer Decoder uses cross-attention to map the encoded features to 3D coordinates, generating a point cloud that represents the spatial structure of the environment.

\subsection{Temporal Encoding}

The reshaped CSI data is processed by the Temporal Encoder, which extracts temporal patterns using a convolution across the time slices for each antenna-subcarrier pair. For each pair $i$, the input CSI data over time is represented as $ \mathbf{x}_i \in \mathbb{R}^{T \times 2} $, where $T$ denotes the number of time slices, and 2 corresponds to amplitude and phase. A one-dimensional convolution is applied to $ \mathbf{x}_i $, resulting in a temporal feature vector $ \mathbf{h}_i \in \mathbb{R}^E $, where $E$ is the embedding dimension. This process, expressed as $h_i = \text{Conv1D}(x_i)$, captures temporal variations in amplitude and phase, reflecting environmental changes. These temporal features $ \mathbf{h}_i $ are then passed to the transformer to learn both temporal and spatial dependencies.

\subsection{Positional Encoding}

To incorporate the spatial structure of the input, we introduce positional encodings for both antennas and subcarriers. These encodings are derived from the indices of the antennas and subcarriers, which provide spatial context to the model. Two learned embedding matrices are used: one for antennas and one for subcarriers. The positional encoding for each pair is computed as:

\begin{equation}
\begin{aligned}
\mathbf{p}_a &= \text{Embedding}_a(a) \\
\mathbf{p}_s &= \text{Embedding}_s(s) 
\end{aligned}
\end{equation}

where $\mathbf{p}_a$ and $\mathbf{p}_s$ represent the positional embeddings for antennas and subcarriers, respectively. The combined encoding $\mathbf{p}_{a+s}$ is added to the temporal features, ensuring that the model preserves both the spatial arrangement and temporal dynamics of the CSI data.

\subsection{Transformer Encoder and Decoder}

The encoded data is then passed through the Transformer Encoder, which applies multi-head attention to capture interactions between the antenna-subcarrier pairs over time. The attention mechanism computes a weighted sum of the input features based on learned attention scores. The self-attention mechanism is given by:

\vspace{-3mm}
\begin{equation}
\text{Attention}(Q, K, V) = \text{softmax}\left( \frac{QK^T}{\sqrt{d_k}} \right) V
\end{equation}

where $Q$, $K$, and $V$ represent the query, key, and value matrices, and $d_k$ is the dimensionality of the key vectors. This attention mechanism allows the model to focus on different aspects of the input simultaneously, capturing complex dependencies across the time and feature dimensions.

The Transformer Decoder generates the 3D point cloud by processing the output from the encoder and mapping it to 3D coordinates. The decoder uses cross-attention between the encoded features and learned point queries to refine the spatial representation. The final output is a set of 3D points, $\mathbf{p}_i \in \mathbb{R}^3$, where each point corresponds to $(x, y, z)$ coordinates in the reconstructed point cloud. The projection is defined as $\mathbf{p}_i = \mathbf{W} \mathbf{h}_i + \mathbf{b}$, where $\mathbf{W}$ is the weight matrix and $\mathbf{b}$ is the bias term. While a PointNet \cite{qi2017pointnet} module was also tested for the output projection, results were very similar to those achieved with the linear projection. The final output shape is $[batch_size, num_points, 3]$, where each point corresponds to a spatial coordinate in the point cloud.

\subsection{Loss Function}
\label{sec:loss_function}

The CSI2PC model uses a combined loss function that includes \textbf{Chamfer Loss} for geometric accuracy and the \textbf{Feature Transform Regularizer} to ensure the transformation matrix stays close to orthogonal. Chamfer Loss function measures the geometric similarity between predicted and target point clouds, ensuring the generated 3D structures match the ground truth \cite{lin2024chamferloss}. This formulation ensures that each point in the predicted point cloud is close to the nearest point in the target point cloud and vice versa.

To maintain the stability of the learned transformation, we incorporate the \textbf{Feature Transform Regularizer}. This regularization ensures that the transformation matrix $\mathbf{T} \in \mathbb{R}^{K \times K}$, where $K$ is the feature space dimension, remains close to an orthogonal matrix. By penalizing deviations from orthogonality, the regularizer prevents excessive distortion of the feature space during transformation, which is essential for maintaining high-quality point cloud reconstructions. The regularizer is defined as:

\begin{equation}
\mathcal{L}_{\text{FeatureTransform}}(\mathbf{T}) = \frac{1}{K^2} \|\mathbf{T} \mathbf{T}^\top - \mathbf{I}\|_F^2
\end{equation}

where $\mathbf{I} \in \mathbb{R}^{K \times K}$ is the identity matrix and $\|\cdot\|_F$ is the Frobenius norm. This loss term ensures that $\mathbf{T} \mathbf{T}^\top$ remains close to the identity matrix, promoting orthogonality and improving feature alignment, which is critical in our context for avoiding degenerate transformations that could lead to poor point cloud estimation.

The total loss function used to train the model combines the Chamfer Loss and the Feature Transform Regularizer, weighted by a hyperparameter $\lambda$, which controls the strength of the regularization:

\begin{equation}
\mathcal{L}_{\text{total}} = \mathcal{L}_{\text{Chamfer}} + \lambda \mathcal{L}_{\text{FeatureTransform}}
\end{equation}

In this context, this combined loss ensures accurate 3D reconstructions while maintaining stable feature transformations. Chamfer Loss promotes geometric accuracy, and the regularizer preserves feature alignment during training.

\section{Experimental Setup}

\subsection{Benchmark database}
The dataset used in this study is the MM-Fi dataset \cite{yang2024mmfi}, designed for multi-modal human sensing tasks, such as 2D/3D human pose estimation and action recognition. It contains data from 40 participants (11 females, 29 males) who performed 27 actions including daily activities and rehabilitation exercises. The dataset includes data from six sensor modalities: RGB, infra-red, depth, LiDAR, mmWave radar, and WiFi CSI. The data was collected in four different environments, which involved two room configurations (both 8.5m x 7.8m), with 10 participants in each environment. The actions cover 14 daily activities (e.g., chest expansion, waving hands, picking up objects) and 13 rehabilitation exercises (e.g., squats, lunges, limb extensions). These actions reflect human motions relevant to healthcare and home automation applications.

During the data collection, the sensors were positioned 3 meters from participants, with the WiFi transmitter placed 0.75 meters from the subject. A Robot Operating System (ROS) synchronized all sensor streams, maintaining a maximum time offset of 25ms between modalities. Data was aligned to a uniform sampling rate of 10Hz by down-sampling higher frequency sensors like LiDAR and mmWave radar. \textbf{WiFi CSI Data:} WiFi Channel State Information (CSI) was collected using TP-Link N750 Access Points, which operate at a 5GHz frequency with a 40MHz bandwidth. The system uses 1 transmitter and 3 receiver antennas, collecting CSI data across 114 subcarriers for each antenna pair. The raw CSI data is captured at a high sampling rate of 1000Hz, then averaged to 100Hz using a sliding window method to smooth the data. To capture detailed WiFi CSI data, researchers developed customized OpenWrt firmware for TP-Link N750 devices using the Atheros CSI Tool. This tool extracts detailed PHY information, including CSI, from Atheros 802.11n NICs. Captured CSI data is a complex matrix representing channel state for each subcarrier, enabling analysis of spatial and temporal patterns affected by human movement. For 40MHz bandwidth, the CSI matrix is 3 × 1 × 114. \textbf{LiDAR Data:} The Ouster OS1 32-channel LiDAR captured 3D point clouds at 10Hz with 32 vertical beams and a vertical angular resolution of ±0.7 to 5 cm, generating up to 1,310,720 points per second. Each point in the cloud represents spatial coordinates (x, y, z). \textbf{mmWave Radar Data:} The Texas Instruments IWR6843AOP mmWave radar, operating at 60-64GHz, captured 30 frames per second. Point clouds contained (x, y, z), Doppler velocity, and intensity data.\textbf{RGB-D Camera Data:} The Intel RealSense D435 stereo depth camera captured RGB-D frames at 30Hz with a resolution of 1920x1080 for RGB and 1280x720 for depth. An IR projector enhanced depth sensing in low-light conditions.

Our study focuses on three modalities: LiDAR, WiFi CSI, and depth camera data. Unlike previous work using human key points for pose estimation, we estimate 3D point clouds directly from these sensors. WiFi CSI detects spatio-temporal patterns such as multipath and phase variation, providing insights into human movement dynamics and spatial indoor geometry.

\subsection{Protocols}

As we did not aim to use the dataset for its originally intended purpose, we defined two protocols to evaluate our architecture. \textbf{1) Subject-based Split Protocol:} Similar as proposed on \cite{yang2024mmfi}, the model is trained using data from 32 subjects and tested on data from 8 different subjects (20\% of the total). Among the training subjects, 2 were set aside for validation, while the remaining 30 were used for training. Importantly, the model is not exposed to any data from the testing subjects during training, ensuring that the evaluation is performed on previously unseen individuals. \textbf{2) Room-based Split Protocol:} focusing on spatial generalization, the model is trained and validated using data collected in Room 1 (environments E01 and E02) and tested on data from Room 2 (environments E03 and E04). This approach aims to assess the ability of the model to generalize across different room configurations, determining whether it can adapt from one environment to another or similar spaces.

\subsection{Performance Metrics}

Comparing point clouds is challenging due to differences in density, noise, and alignment. To complement the qualitative analysis, we use the Iterative Closest Point (ICP) algorithm \cite{besl1992method} to assess alignment between predicted and ground truth point clouds. This yields two metrics: ICP Fitness, which measures the proportion of points in the predicted cloud aligning with the ground truth within a threshold, and ICP RMSE, which calculates the average Euclidean distance between corresponding points. The alignment minimizes the distance between points using the following ICP registration:

\begin{equation}
T = \arg \min_T \sum_{i} \| p_i - T(q_i) \|^2
\end{equation}

where $p_i$ are ground truth points, $q_i$ are predicted points, and $T$ is the transformation matrix. Higher ICP Fitness indicates better alignment, while lower RMSE shows greater spatial accuracy, providing an objective measure of model performance in generating 3D reconstructions from WiFi-CSI data.





\subsection{Implementation details}

The input data for the CSI2PC model consists of CSI data vectors of shape $[3, 114, 2, 10]$, corresponding to 3 antennas, 114 subcarriers, 2 channels (amplitude and phase), and 10 time slices. Each CSI input is paired with a ground truth 3D point cloud of dimension $[1200, 3]$, representing 1200 points in 3D space. The model is trained for 50 epochs using the \textit{NAdam} optimizer with a learning rate of $1 \times 10^{-4}$. The model configuration includes an embedding dimension of 512, with 4 attention heads, 4 encoder layers, and 4 decoder layers. The number of output points is set to 1200 to match the ground truth, and the CSI data is processed using 3 antennas, 114 subcarriers, and 10 time slices per sample. The learning rate scheduler is configured with a step size of 10 epochs and a decay factor $\gamma$ of 0.5. This adjusts the learning rate after every 10 epochs to improve convergence. The implementation is based on PyTorch, using Python 3.10. Model training was performed on Mahti supercomputer (CSC Finnish Supercomputer Center), which is a BullSequana XH2000 system. The model training utilized 2 NVIDIA$^\circledR$ Ampere A100 GPUs (40GB). Each model required approximately 16 hours of training. For testing, a high-performance workstation was used, equipped with an AMD$^\circledR$ Ryzen 3700X 8-core processor, 64GB of RAM, and two NVIDIA GeForce$^\circledR$ RTX 2080 GPUs.

%
%
%
%


\section{Experimental Results}

We primarily evaluate qualitative results to visually assess the effectiveness of using CSI-WiFi data for estimating 3D indoor point clouds, given the novelty of the approach. These visual evaluations provide insight into how well the model captures spatial structures. In addition to qualitative analysis, we also use quantitative metrics, laying the groundwork for further detailed analysis and benchmarking in future work.

\subsection{Qualitative Results}
\begin{figure*}[ht!]
 \begin{center}
   \includegraphics*[width=0.89\textwidth]{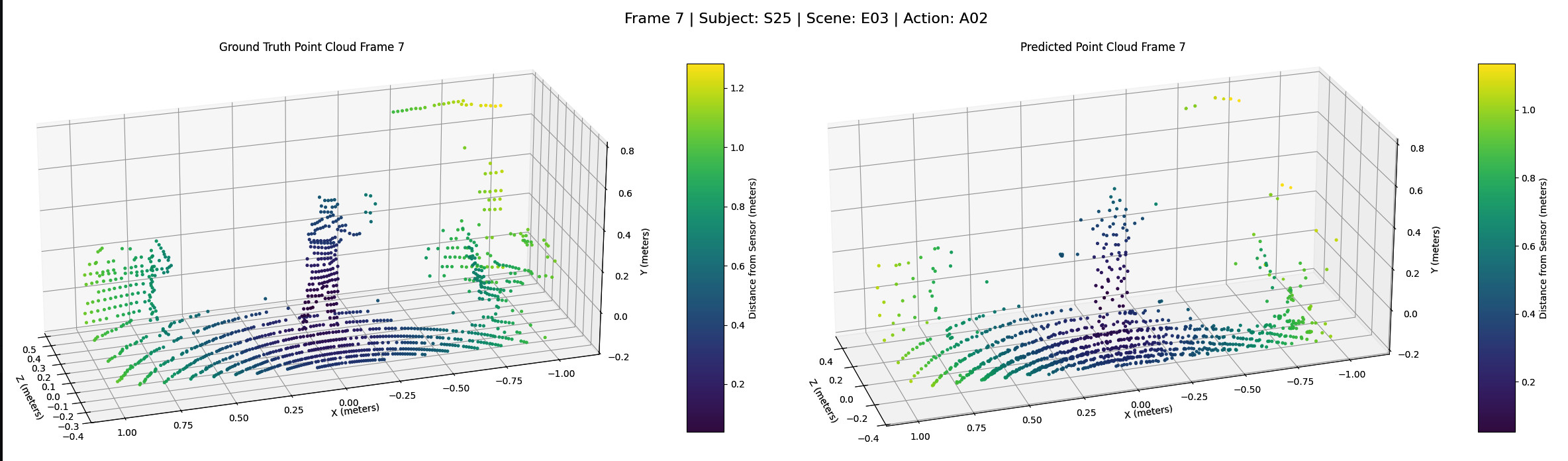}
 \end{center}
 \caption{Comparison between the Ground Truth Point Cloud (left) and the Predicted Point Cloud (right) for Frame 7 of Subject S25 and Action A02. The ground truth shows the actual 3D structure, while the prediction captures general patterns with some differences in object positioning and density.}
 \label{fig:subjectProtocolResults}
\end{figure*}

Figure \ref{fig:subjectProtocolResults} shows a qualitative comparison between the ground truth 3D point cloud (left) obtained from the LiDAR sensor and the estimated point cloud (right) generated by the model using CSI-WiFi data. This particular result corresponds to the subject-based split protocol, where the subject shown in the image was not seen during training. In the estimated point cloud, the position of the subject in the room is clearly identifiable, with approximate distances and dimensions similar to those in the original ground truth image. Although the resolution and number of points in critical areas are lower than the ground truth, the generated point cloud retains enough detail to distinguish both the human subject and surrounding furniture. These results highlight the model's ability to capture key spatial features despite lower point density.

\begin{figure}[ht!]
 \begin{center}
   \includegraphics*[width=0.49\textwidth]{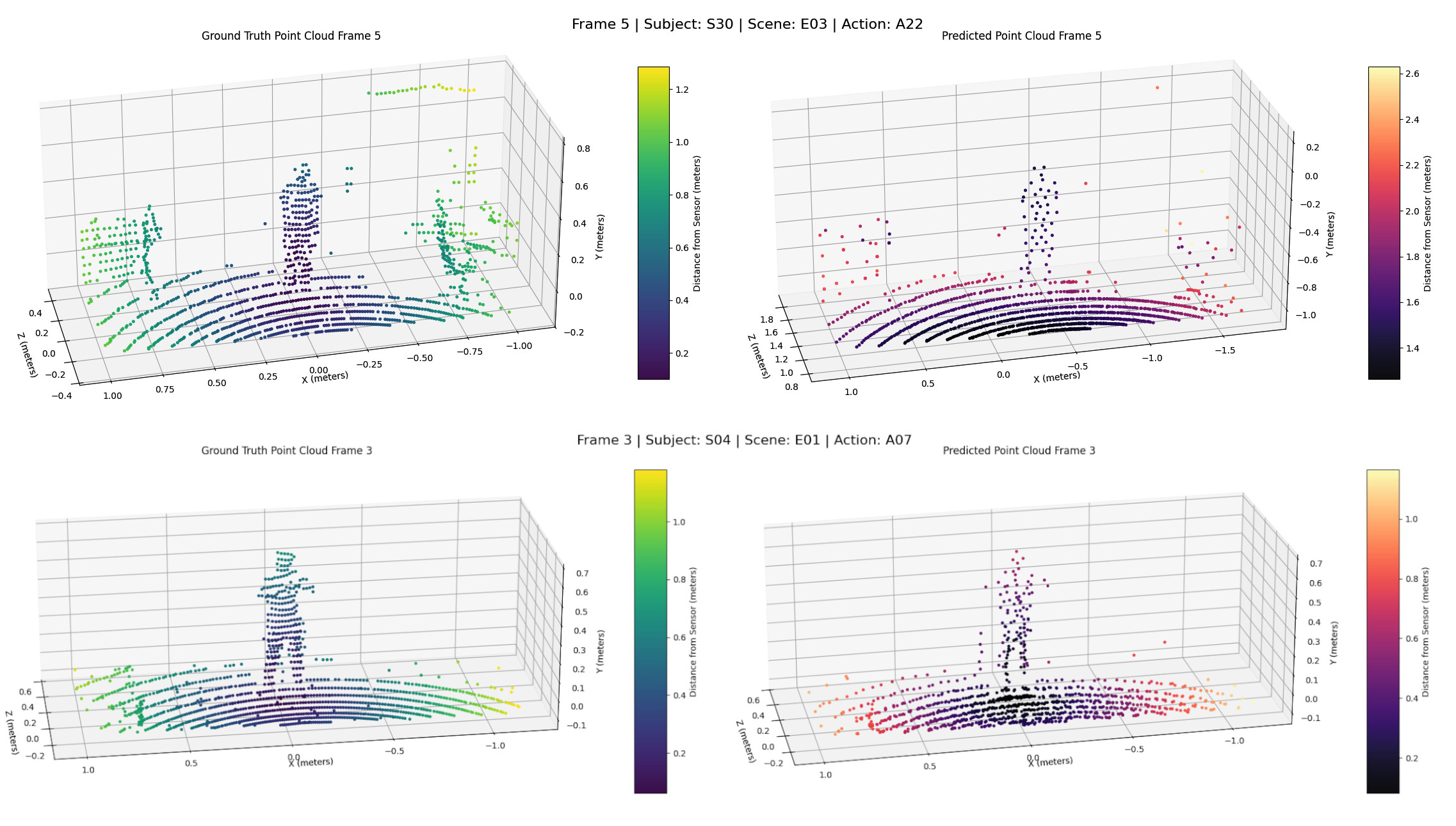}
 \end{center}
 \caption{Predictions using the Room-Split Protocol. The upper part shows Subject S30 in Room 2 (E03), predicted by a model trained on Room 1 data. The lower part shows Subject S04 in Room 1 (E01), predicted by a model trained on Room 2 data.}
 \label{fig:RoomProtocolResults}
\end{figure}

Figure \ref{fig:RoomProtocolResults} shows the predictions using the room-split protocol. In the upper part of the figure, the model predicts Subject S30 in Room 2 (E03), using a model trained exclusively with data from subjects in Room 1. In the lower part of the figure, Subject S04 performing an action in Room 1 (E01) is predicted by a model trained only with subjects from Room 2. This setup corresponds to the Room-based Split Protocol, which focuses on spatial generalization. The model is trained and validated using data collected in Room 1 (environments E01 and E02) and tested on data from Room 2 (environments E03 and E04). This protocol assesses the model's ability to generalize across different room configurations, testing whether it can adapt to changes in environment or similar spaces. The results suggest that while the model retains a degree of spatial accuracy, performance can vary depending on room-specific features and signal characteristics.

However, some limitations are evident, particularly in terms of depth estimation and vertical angle coverage and resolution. These limitations arise from the proximity between the transmitter (TX) and receiver (RX) in the setup, as well as constraints related to the LiDAR sensor itself. Despite these challenges, the estimated point cloud provides sufficient information for approximate spatial reconstruction.

\subsection{Quantitative Results}

We evaluate the performance of the CSI2PC model in generating 3D point clouds using ICP Fitness, ICP RMSE, and time consumption on both CPU and GPU. These metrics assess spatial alignment accuracy and computational efficiency. As no direct comparison methods are available, we present results from both the subject-based and room-based split protocols.

The results are summarized in Table \ref{tab:tableICP}. In the subject-based split protocol, the model achieves higher ICP Fitness and lower RMSE, demonstrating better performance when generalizing across subjects. In contrast, the room-based split protocol shows slightly reduced performance, likely due to the challenges posed by differences in room geometry, which affect CSI signal propagation and spatial adaptation.Our model runs relatively fast on both CPU and GPU, even when processing data in Float32 without any quantization or model optimization strategies, as evidenced by the time consumption metrics.

\begin{table}[ht!]
\setlength{\tabcolsep}{1.2em}
\def\arraystretch{1.1}
\centering
\caption{Quantitative Results of CSI2PC Model using ICP Fitness, RMSE, and Average Time per Estimation}
\label{tab:tableICP}
\begin{tabular}{|l|c|c|}
\hline \textbf{Metric} & \textbf{Subject-based Split} & \textbf{Room-based Split} \\
\hline ICP Fitness & 0.6358 & 0.6127 \\
\hline ICP RMSE (m.) & 0.0103  & 0.0104 \\
\hline CPU (ms.) & 137.27 & 135.79 \\
\hline GPU (ms.) & 2.31 & 2.30 \\
\hline
\end{tabular}
\end{table}

\section{Conclusion}

This work introduced a transformer-based architecture, CSI2PC, designed to estimate 3D point clouds from WiFi-CSI data. Using the MM-Fi database, the model was trained and evaluated across different subjects and room configurations, demonstrating its capacity to generate approximate 3D reconstructions. Despite the challenges inherent in using WiFi-CSI data, the model effectively captured key spatial features such as human presence, furniture, and room geometry. The evaluation through subject-based and room-based split protocols highlighted the ability of the proposed model to generalize to unseen individuals and adapt to different spatial environments. However, limitations were observed in the depth and vertical angle estimation, which are attributed to the proximity of the transmitter and receiver, as well as constraints in the ground truth data. 

This approach opens new possibilities for joint communication and sensing (JC\&S), where future wireless networks could integrate environmental awareness as a built-in feature of communication systems. The generalization capabilities observed in the subject-based and room-based splits provide a foundation for future work, focusing on improving depth estimation, handling room-specific geometries, and exploring multimodal sensor fusion. This could enhance the accuracy and robustness of WiFi-CSI-based spatial sensing, making it a valuable tool in smart environments, healthcare, and beyond.



\section*{Acknowledgment}
The research was supported by the Research Council of Finland (former Academy of Finland) 6G Flagship Programme (Grant Number: $346208$), the Horizon Europe CONVERGE project (Grant 1010948), Business Finland WISEC project (Grant 3630/31/2024). The authors wish to acknowledge CSC-IT Center for Science, Finland, for computational resources.

\bibliographystyle{IEEEtran}
\bibliography{references}

\end{document}